# Quasiparticle relaxation dynamics in Hg-1223 studied by femtosecond time-resolved optical spectroscopy.

J. Demsar[a], R. Hudej[a], J. Karpinski[b], V.V. Kabanov[a] and D. Mihailovic[a]

[a]*Jozef Stefan Institute, Jamova 39, 1001 Ljubljana, Slovenia*
[b]*Festkorperphysik, ETH CH-8093 Zurich, Switzerland*

Following recent progress in understanding the relaxation dynamics of photoexcited carriers in materials exhibiting a small gap in the low-energy excitation spectrum we have performed pump-probe measurements on near optimally doped Hg-1223. We show that the behavior is very similar as in optimally-doped YBCO, where the data can be interpreted with the coexisting presence of two energy gaps: normal state T-independent pseudogap and a mean-field-like collective gap, associated with intrinsic spatially inhomogeneous ground state. An important difference between the two compounds is found in the low temperature relaxation time, which in Hg-1223 is found to be strongly temperature dependent.

In femtosecond pump-probe experiments, an ultrashort laser *pump* pulse first excites electron-hole pairs via an interband transition in the material. In a process, which is similar in most materials including metals, semiconductors and superconductors [1], these hot carriers very rapidly release their energy via *e-e* and *e-ph* collisions reaching states near the Fermi energy within 10 - 100 fs. The superconducting gap inhibits further relaxation and photoexcited carriers accumulate above the gap [1]. The bottleneck causes a transient change in reflectivity $\Delta R/R$, with the amplitude of the reflectivity transient being proportional to the photoexcited carrier density [1]. Because the final relaxation step across the gap is strongly suppressed, the quasiparticles (QP) together with high frequency phonons (with $\omega > 2\Delta$) form a near-steady state distribution, with the QP recombination dynamics of this system being governed by the emission and reabsorption of high frequency phonons. As the gap closes (in case of T-dependent BCS-like gap), more and more high frequency phonons have the energy to excite Cooper pairs, therefore the relaxation time $\tau_R$ is expected to show a divergence [1] as $\tau_R \sim 1/\Delta(T)$. From the analysis of amplitude and relaxation time of the *ps* reflectivity transient one can obtain the magnitude and the T-dependence of the low energy gap [1].

Here we report the first measurements of QP

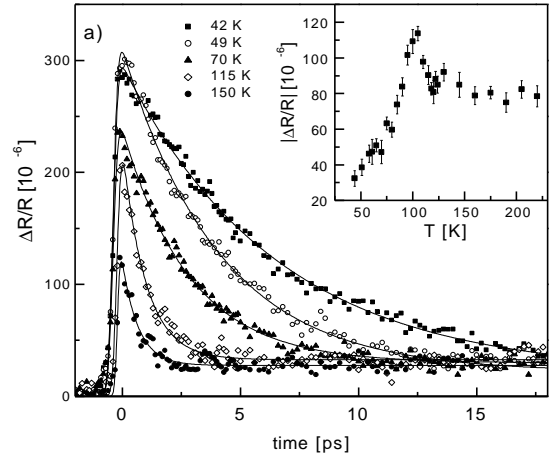

Figure 1. a) $\Delta R/R$ in Hg-1223 at various temperatures below and above $T_c$. In all traces a T-dependent background is subtracted. (Inset: T-dependence of background amplitude, which shows similar T-dependence as in near-optimally doped YBCO [7]).

dynamics on $HgBa_2Ca_2Cu_3O_{8+x}$ (Hg-1223) with $T_c$=120 K, as determined by AC susceptibility measurements. The experiments were performed at typical pump pulse fluence $E_0$=1.3 µJ/cm$^2$; all the experimental details are described elsewhere [2]. In particular, we focus on the temperature dependence of QP relaxation time, which shows some differences with respect to YBCO - especially at low temperatures.

In Fig.1 a) the photoinduced reflectivity transients taken at different temperatures below and above $T_c$ are plotted. All seem to be well reproduced by single exponential decay (solid lines). Both the relaxation time and $\Delta R/R$ amplitude are strongly temperature dependent. The T-dependence of $\Delta R/R$ amplitude [Fig. 2 a)] is very similar to the observed behavior in optimally doped and overdoped YBCO [3] - it is constant at low temperatures, showing a rapid drop as $T_c$ is approached. Note that above $T_c$, $\Delta R/R$ amplitude asymptotically decreases, similarly as in overdoped YBCO, suggesting the presence of a normal state gap above $T_c$ in Hg-1223.

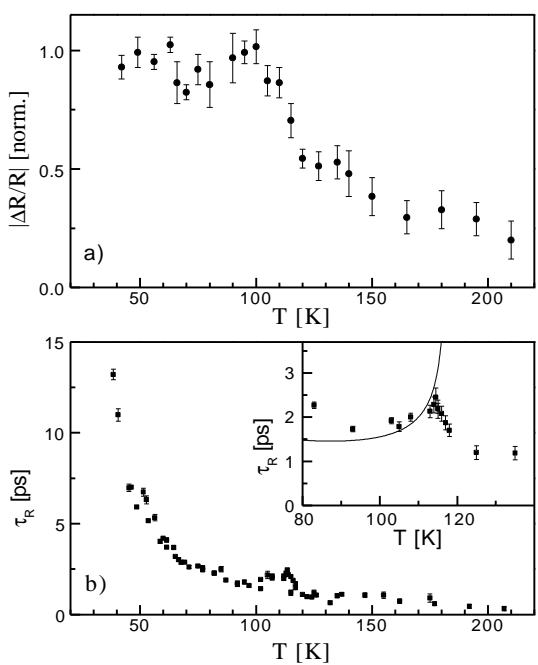

Figure 2: a) T-dependence of the picosecond component amplitude. b) T-dependence of the QP relaxation time. The divergence of $\tau_R$ at $T_c$ is shown in inset.

The T-dependence of the QP relaxation time determined by single exponential fit to the data shows substantially different behavior to YBCO [Fig. 2 b)]. At low temperatures, the relaxation time rapidly increases upon decreasing the temperature.

Similar behavior has been recently observed at low temperatures in BiSCO-2212, Tl-2201 [4] and NCCO [5]. In NCCO a strong increase in the QP relaxation time is observed already at ~ 80 K (far above $T_c$), suggesting that the origin of this low-temperature decrease may not be connected to the superconducting state. As $T_c$ is approached from below, $\tau_R$ shows a divergence [see inset to Fig. 2 b)] as observed already in YBCO, Tl-2223 [3,6] and also in NCCO [5]. This is, through the relation $\tau_R \propto 1/\Delta(T)$, associated with the closure of the isotropic collective mean-field-like gap [1]. Above $T_c$ the relaxation time drops to ~ 1 *ps*, which is much larger than usual metallic relaxation [1] and is attributed to the presence of the normal state gap.

In YBCO two distinct relaxation components below $T_c$ (one present also above $T_c$ with T-independent $\tau$, and the other with divergent $\tau_R$ at $T_c$ [3,6]) suggest the *co-existing* presence of two distinct energy gaps: a T-independent pseudogap $\Delta^p$ and a mean-field-like T-dependent gap $\Delta_c(T)$ [3]. Two distinct relaxation components with opposite signs were observed also on BiSCO-2212 and Tl-2201 [4], suggesting that the two component behavior is quite general in high-$T_c$ superconductors near optimal doping. On Hg-1223, the relaxation seems to be well reproduced by single exponential decay, however the presence of a signal above $T_c$ together with an asymptotic decrease in $\Delta R/R$ amplitude at high temperatures suggests similar two-component behavior as for YBCO, with the two relaxation times too close to be resolved.